\documentclass{article}

\usepackage[english]{babel}

\usepackage[letterpaper,top=2cm,bottom=2cm,left=3cm,right=3cm,marginparwidth=1.75cm]{geometry}

\usepackage{amsmath}
\usepackage{graphicx}
\usepackage{amssymb}
\usepackage[colorlinks=true, allcolors=blue]{hyperref}
\usepackage[algo2e, ruled,vlined,linesnumbered]{algorithm2e}
\usepackage{subcaption}
\usepackage{xcolor,colortbl}
\usepackage{makecell}
\usepackage{hyperref}

\usepackage{appendix}
\usepackage{caption}
\usepackage{multirow}
\usepackage{authblk}

\DeclareMathOperator*{\argmax}{arg\,max}
\DeclareMathOperator*{\argmin}{arg\,min}

\newcommand{\ourmethod}{Relations-Maximization Method\xspace}
\newcommand{\bx}{\mathbf{x}}
\newcommand{\bw}{\mathbf{w}}

\title{Network Based Approach to Gene
Prioritization at Genome-Wide
Association Study Loci}

\author[1,2,3]{Leonardo Martini}
\author[4]{Adriano Fazzone}
\author[2,3]{Michele Gentili}
\author[1]{Luca Becchetti}
\author[2,3]{Brian Hobbs}

\affil[1]{Department of Computer, Control, and Management Engineering Antonio Ruberti, Sapienza University of Rome, Rome, Italy}
\affil[2]{Channing Division of Network Medicine, Brigham and Women’s Hospital, Boston, MA}
\affil[3]{Harvard Medical School, Boston, MA}
\affil[4]{CENTAI Institute, Turin, Italy}

\begin{document}
\maketitle

\begin{abstract}
\noindent \textbf{Motivation:} Genome-wide association studies (GWAS) have successfully identified thousands of genetic risk loci for complex traits and diseases. Most of these GWAS loci lie in regulatory regions of the genome and the gene through which each GWAS risk locus exerts its effects is not always clear. Many computational methods utilizing biological data sources have been proposed to identify putative casual genes at GWAS loci; however, these methods can be improved upon.\\
\textbf{Results:} We present the \ourmethod, a dense module searching method to identify putative causal genes at GWAS loci through the generation of candidate sub-networks derived by integrating association signals from GWAS data into the gene co-regulation network. We employ our method in a chronic obstructive pulmonary disease GWAS. We perform 
an extensive, comparative study of \ourmethod's performance against 
well-established baselines.\\
\textbf{Availability:} The code and the datasets used in this manuscript are available at the following website:\\\end{abstract}

\section{Introduction}

Genome-wide associations studies (GWAS) have successfully identified thousands of genetic risk loci for complex traits and diseases; however, translation of the risk loci to disease-causing genes and biologic pathways is challenging. Often, researchers approach genetic risk loci one at a time using a series of bioinformatics methods. A typical approach involves assessing if the top variant at each disease risk locus is an expression quantitative trait locus (eQTL) - meaning that the disease risk variant is associated with gene expression levels - in disease-relevant tissues\cite{RN5520}. Statistical colocalization of GWAS  and eQTL association signals can further increase confidence in the most likely effector gene at each GWAS risk locus\cite{RN6099}. However, considering each locus in isolation does not leverage the fact that genetic risk for complex diseases is additive across many loci\cite{RN8786, RN8787, RN8336}. The genetic risk to most diseases is likely distributed across a large number of biologic processes\cite{RN8788} with a network of molecular interactions contributing to disease biology\cite{RN107}.
\newline

\noindent 
For this reason, several tools have been designed to identify disease-causing genes starting from GWAS studies. Some leverage network topology to handle the complex interplay of biomolecules that lead to disease, we will refer as Network-based approaches\cite{furlong2013human,leiserson2013network}. While other methods functionally annotate GWAS findings and prioritize the most likely causal genes using several biological data sources\cite{climente2021boosting}. They are referred as integrative approaches. 
\newline

\noindent Network-based approaches first build a weighted network of associations between gene and risk locus, and then they find a subset of highly scored genes and locus that are  functionally-related and densely connected.
LEAN\cite{gwinner2017network} predicts a "star" disease module in which a gene and its direct interactors are associated with the disease. DmGWAS\cite{jia2011dmgwas}, EWdmGWAS\cite{wang2015ew_dmgwas} and Heinz\cite{dittrich2008identifying} find a  disease module consisting of genes with a high association score. HotNet2\cite{leiserson2015pan}, and SigMod\cite{liu2017sigmod} find disease modules of densely connected genes but not necessarily high scoring. SCoES\cite{azencott2013efficient} leverages the SNP network, consisting of SNPs that interact if there exists a protein-protein interaction between their associated genes, to discover a disease module. DOMINO\cite{levi2021domino} leverages differential gene expression data to detect active network modules that are statistically significant in GO terms that do not appear to be enriched in random permutations.

\noindent
Integrative approaches, instead, combine different biological information to prioritize candidate disease genes. DEPICT\cite{fehrmann2015gene} leverages co-regulation of gene expression to predict gene function based on a guilt-by-association procedure and prioritizing genes at different risk loci that have similar predicted functions. FUMA \cite{watanabe2017functional} is an integrative platform that combines positional mapping, expression quantitative trait loci (eQTL) mapping, chromatin interaction mapping, and MAGMA\cite{de2015magma} gene analysis to predict candidate disease genes. MendelVar\cite{sobczyk2021mendelvar} prioritize candidate genes by mapping GWAS traits to known phenotypically relevant Mendelian disease genes near loci.

\noindent

\paragraph{Our Contribution.} In the following manuscript, we present the \textbf{\ourmethod}, a network-based approach that leverages co-regulatory networks and biological annotations to find the most likely functional set of genes associated to the risk loci of a complex disease. It is built on the hypothesis that disease-relevant genes  across GWAS loci share biological processes, pathways, and correlated gene expression levels. Thus, given a complex disease with multiple risk loci, the proposed algorithm prioritizes a set of genes, that belong to different loci, so that the number of co-regulatory interactions, weighted by the number of shared biological annotations, is maximized.

\noindent
To demonstrate the application of our method, we chose chronic obstructive pulmonary disease (COPD) as our complex genetic disease of interest. COPD is a debilitating respiratory disease affecting millions of individuals across the globe and is estimated to be the third leading cause of death worldwide.\cite{RN8388} The diagnosis and severity of COPD are defined by summary measures of lung function; however, COPD is clinically heterogeneous, and individuals with similar lung function impairment may have vastly different appearances of their lung parenchyma on computed tomography (CT) imaging.\cite{RN5408} The clinical heterogeneity of COPD also likely contributes to the difficulty of developing broadly effective pharmacologic therapies for reducing COPD mortality. Only recently, in a trial of >10,000 individuals, inhaled therapy with steroids was shown to have a mortality benefit over inhaled therapy with bronchodilator medications alone\cite{RN8865}. These inhaled therapies are the mainstay of COPD treatment, though multiple previous trials of similar inhaled therapy regimens had failed to demonstrate a mortality benefit in COPD patients. New approaches to COPD therapy are needed. Drug targets with human genetics support, as from GWAS, are more likely to be successful through the rigorous clinical investigations required for drug approval\cite{RN8868, RN8869}. The genetic data we are employing for COPD is from a 2019 COPD GWAS, which identified up to 82 genetic loci associated with the risk of COPD\cite{RN7295}.

\section{Methods}

\subsection{Data}\label{subse:data}

\subsubsection{Biological data sources}

\paragraph{GWAS Summary Statistics:} We downloaded COPD GWAS Summary statistics from\cite{sakornsakolpat2019genetic}. We downloaded Breast Cancer GWAS Statistics from the GWAS catalog.

\paragraph{Gene-term associations.}
We used two different sources of biological information: 
\begin{itemize}
    
    \item Gene Ontology 
    Consortium\footnote{
    \url{http://geneontology.org/}
    }: For each gene 
    we downloaded its associated terms (cellular components, molecular 
    pathways and biological processes), filtering out terms that not 
    been manually reviewed (IEA evidence code).
    
    \item Reactome\cite{fabregat2016reactome}: this is an open-source, curated and peer-reviewed database that associates genes with their biological pathways.
\end{itemize}

\paragraph{Drug-target associations.} We downloaded drug-target associations from the Drug Repurposing Hub Database\cite{corsello2017drug}.

\paragraph{Gene Ontology Graph.} We downloaded the Gene Ontology Graph 
from the Gene Ontology Consortium. This is  a network in which each 
node is a gene ontology annotation/term, while edges represent 
relationships between annotations/terms. 
This network is loosely hierarchical, with \textit{child} nodes more specialized 
than their \textit{parent} nodes. 
Unlike a strict hierarchy, however, a node 
may have more than one parent. The relationships we used to filter the \textit{GO Graph} are:
\begin{itemize}

    \item \textbf{``is a'':} This relation forms the basic structure of GO. If a pair $(a,b)$ has this relationship code, then $a$ is a subtype of $b.$
    
    \item \textbf{``part of'':} This relation is used to represent 
    part-whole relationships. This relation has a specific meaning in 
    GO: it is associated to a pair $(a,b)$ if $b$ is necessarily part 
    of $a$. That is, the presence of the $b$ implies the presence of $a$, 
    though the converse may not be true.
\end{itemize}

\paragraph{Co-Regulation Network.} we downloaded co-regulation network from in Fehrmann et al~.\cite{fehrmann2015gene} (and \url{www.genenetwork.nl}), filtering out not statistically significant edges (i.e. P-value $> 10^{-2})$. 

%%%
\subsection{Data Pre-processing}\label{subse:preproc}

\subsubsection{Construction of risk locus - gene bipartite network}
\label{subsec:genelocusassociation}

Starting from a set of SNPs in a GWAS study, we used \textit{bed tool} to select overlapping genes within a \textit{2Mb} window around the SNPs selected by the GWAS study. Then, we assigned a set of genes to each risk locus as follows: a gene is associated to a locus if the index SNP resides within the gene, or if the gene overlaps with a $2Mb$ window around the index SNP; whenever a gene overlaps with multiple loci, it is assigned to the nearest index SNP. In this way, i) each gene is associated with exactly one GWAS risk locus and ii) 
each risk locus corresponds to a unique set of genes. To assign each gene to its nearest risk locus,  we first define the distance between risk locus $l$ (marked by the associated SNPs) and the genetic position of a gene (i.e its starting and ending positions $[g_{start}, g_{end}]$) as:

 \begin{equation}
    D(g,l) =
    \begin{cases}
        \min{\left(| g_{start} - l |,| g_{end} - l |\right)}, & \text{\textit{if} } \left(g \in C \wedge l \in C \right); \\
     \infty,       & \text{\textit{ otherwise}}.
    \end{cases}
  \end{equation}
Where $C$ is the Chromosome.
Secondly, we assign each gene to the risk locus that minimize the distance (i.e. $ \argmin_{l \in L} D(g,l)$ where L is the set of COPD's risk loci). We define positions in the human genome according to genome build \textbf{GRCh37}. 

\subsubsection{Filtering the Co-Regulation Network}
\label{subsec:coregnet}

To model the effects of correlations between genes assigned statistically GWAS risk loci, we combined the co-regulation network and the risk locus - gene associations bipartite graph discussed in section \ref{subsec:genelocusassociation}
to derive a sub-network, deleting those edges that link genes assigned to the same risk locus and deleting genes not associated to any locus. Then, we assign weights to the each edge of this filtered co-regulation network, using biological similarity among genes to this purpose (more details are given in section \ref{subse:preproc_resnik}).

\subsubsection{Weighting co-regulation network using Biological Information}
\label{subse:preproc_resnik}
Each edge in the co-regulation network is assigned a weight that 
depends on the biological similarity between the genes corresponding to 
the edge's endpoints. To do this, we leverage biological processes 
downloaded from Gene Ontology Consortium~\cite{gene2004gene}. 
Similarity between two genes $u$ and $v$ is defined with respect to the 
GO annotations associated to $u$ and $v$, To begin, we use 
\textit{Resnik's method}~\cite{resnik1995using} to assign a semantic 
similarity score to a pair $a$ and $b$ of GO annotations. In more detail, Resnik's method quantifies the similarity measure of a 
term pair $(a, b)$ as the information content (IC) of their Most 
Informative Common Ancestor (MICA), namely: 
\begin{equation}\label{eq:mica}
    R_{sim} (a, b ) = \max{\{IC(c) \mid c \in CA(a, b )\}}.
\end{equation}

In \eqref{eq:mica}, $CA(a, b)$ is the set of common ancestor terms 
between the terms $a$ and $b$ in the GO graph.
$IC(a)$ denotes the information content value of term $a$, defined as 
$-log(p(a))$, with $p(a)$ the probability that $a$ occurs in a certain 
corpus of GO annotation, such as the human annotation database. In 
other word, $p(a)$ is the frequency with which one encounters the annotation 
$a$ within the corpus under consideration. This implies that if an 
annotation $a$ is a parent of another annotation $b$, then we 
necessarily have $IC(a) \le IC(b)$ (i.e the Information content is a 
monotonic function). Of course, a gene can be involved in multiple 
biological processes, thus it could be annotated with multiple GO terms.
For this reason, we estimate the semantic similarity between pairs of 
genes that share an edge in the co-regulation network using the Best Match Average (BMA) 
Rule~\cite{azuaje2005ontology}, whereby the similarity measure 
between genes $u$ and $v$ is defined as the mean of the following two 
values: average of the maximum similarity scores between each GO term 
associated to gene $u$ and those annotated to gene $v$, and the average 
of the maximum similarity scores between each GO term associated to gene $v$ 
and those annotated to gene $u$. In formulas:

\begin{equation}\label{eq:bma}
    BMA(u, v) = \frac{1}{2}  \left(\frac{1}{n} \sum_{t \in T_u} Sim\left(t, T_v\right) + \frac{1}{m} \sum_{t \in T_v} Sim\left(t, T_u\right) \right),
\end{equation}

\noindent where $n$ is the number of annotations associated to $u$ (i.e 
$n = |T_u|$), $m$ is the number of annotations associated to $v$ (i.e 
$n = |T_v|$), $T_u$ and $T_v$ are the sets of $u$'s and $v$'s annotations 
respectively, while $Sim(t, T_u)$ is the maximum Resnik's similarity ($R_{sim}$) between term $t$ and 
GO terms associated to gene $u$. More formally:

\begin{equation}\label{eq:sim}
    Sim(t, T_u) = \max_{s \in T_u} {R_{sim}(t,s)}.
\end{equation}

\begin{figure}
\includegraphics[width=\textwidth]{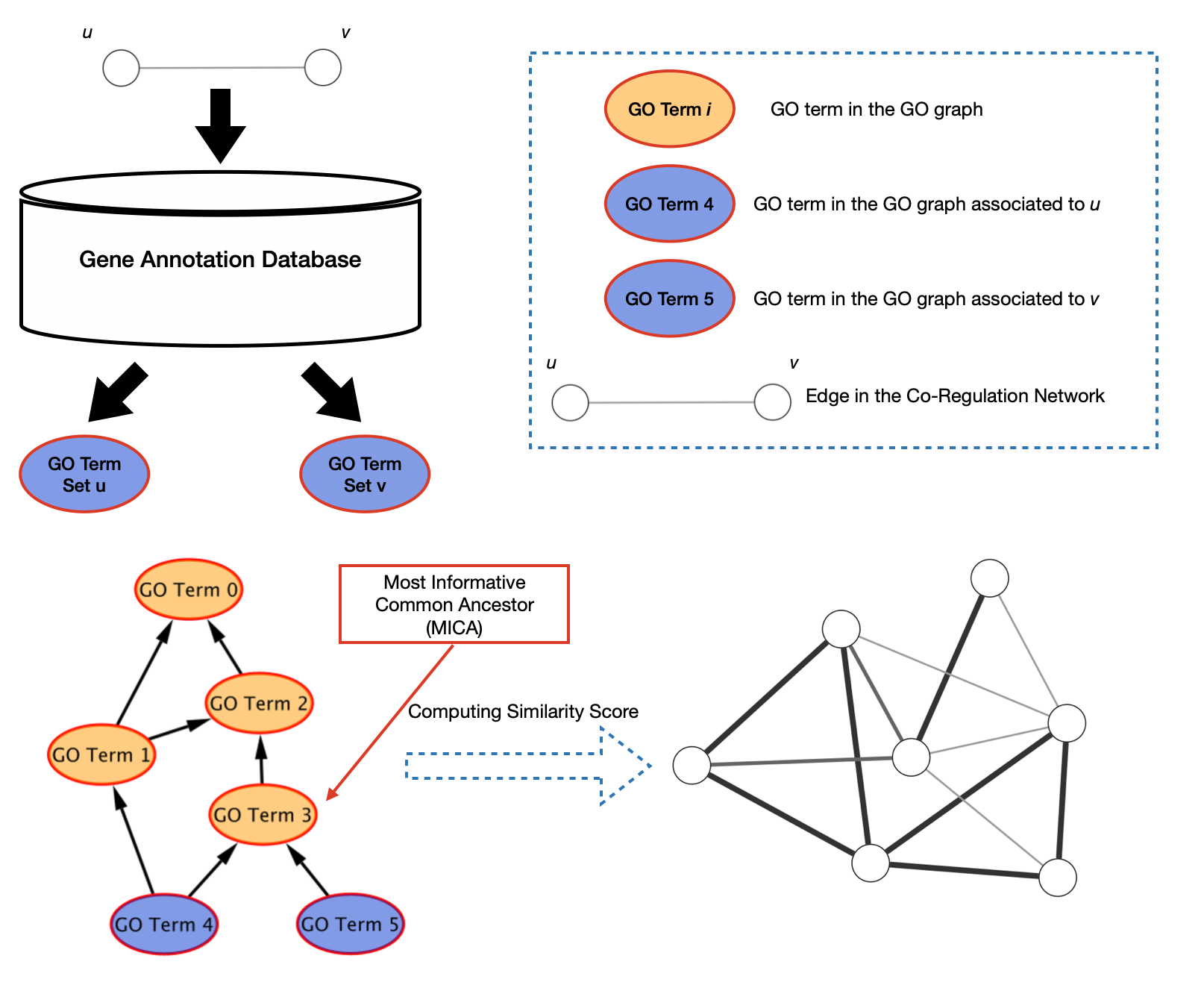}
\caption{Each Link in the Co-Regulation network is weighted using the Go graph, an acyclic network that define the hierarchical relationships between nodes in the graph. Each node represents a biological processes. \textbf{GO Term 3} is the Most Informative Common Ancestor(MICA) and it is used to score the edge $(u,v)$ in the network }
\label{fig:network_preprocessing}
\end{figure}

\noindent Figure \ref{fig:network_preprocessing} describes the steps to weight each link in the network. 
The Final network returned by this biological preprocessing step consists of an undirected weighted graph, 
in which 
nodes are genes associated to the 82 loci of interest, 
and there is an weighted edge between two nodes if they are co-regulated by the same transcription factor and they are biologically similar according to the BMA rule that is used to weight each edge.

\subsection{The \ourmethod}\label{subse:alg}
In this section, we describe our approach to translating GWAS loci to a 
set of likely functional genes. In particular, our aim is twofold: i) 
associating the most likely functional gene to each GWAS locus and ii) 
selecting a subset of genes that collectively share biological processes, pathways,
and correlated gene expression levels, as discussed in the 
introduction. Information about genes' biological similarity is encoded 
in the weighted, undirected network resulting from the biological 
preprocessing described in Section~\ref{subse:preproc_resnik}. 

\paragraph{Formalization as a densest subgraph problem.}
Our input is an undirected, weighted graph $G(V,E, w)$, where each node in 
$V$ is a gene and $(u, v)\in E$ if genes $u$ and $v$ share biological 
features as discussed in Section \ref{subse:preproc_resnik}, while $w: 
E\rightarrow \mathbb{R}^+$ assigns to each edge $(u, v)\in E$ a 
positive weight, reflecting the degree of $u$ and $v$'s biological 
similarity. Moreover, each node has exactly one associated locus. This 
association is described by the function $S \colon V \to \mathcal{L}$, 
for each gene providing the locus it is associated to. On the other hand, each 
locus can have multiple associated genes. For ease of exposition, we 
describe this second relationship as a second function $L \colon 
\mathcal{L} \to \wp(V)$ which, for every locus $\ell$, gives the list 
$L(\ell)$ of the genes that are associated to it.\footnote{Note that 
the use of functions $S$ and $L$ is only to make exposition more 
concise. For example, the set of genes associated to each locus $\ell$ 
can be simply derived from $S$, by taking all genes $v$, such that 
$S(v) = \ell$.}

We next discuss how our two goals i) and ii) mentioned at the beginning 
of this section are mapped into a variant of a well-known optimization 
problem. Our goal is computing a subset $R\subseteq V$ of the nodes 
(i.e., a subset of genes), such that 1) $|R| = |\mathcal{L}|$, i.e., 
the number of genes returned is exactly equal to the number of loci, 2) 
for every $\ell\in\mathcal{L}$, we have exactly one member $v\in R$, 
such that $S(v) = \ell$ and 3) the subgraph induced by $R$ is the 
densest among those satisfying constraints 1) and 2) above. 
Before making point 3) more precise, it is worth noting that 1) and 2) together correspond to achieving goal i) above, returning exactly one associated gene per locus. 
As for point 3), given a  weighted 
graph $G = (V, E, w^+)$ and $R\subseteq V$, the (weighted) subgraph 
induced by $R$ is the graph $G' = (R, E(R), w_R)$, where $E(R)$ 
consists of all edges in $E$ with both end nodes in $R$, i.e., $E = 
\{(u, v)\in E: u, v\in R\}$, while $w_R$ is simply the restriction of 
$w$ to the subset $E(R)$. Given $G = (V, E, w^+)$ and $R\subseteq V$, the 
$\emph{density}$ of the induced subgraph is 
\[
	\delta(R) = \frac{\sum_{e\in E(R)}w(e)}{|R|}.
\]
Intuitively, the higher $\delta(R)$, the higher the average, overall 
weight of the edges incident to each node. In our case, this translates 
to selecting a subset of genes that tend to form a densely connected 
component, with genes of the chosen pool exhibiting above average 
biological similarity. 

Achieving points 1) - 3) above (corresponding to goals i) and ii)) thus 
translates into the following optimization problem. Given $G = (V, E, w^+)$,
%\begin{align}
%	&\max_{R\subseteq V}\frac{\sum_{e\in E(R)}w(e)}{|R|},\label{eq:opt}\\ 
%	&|R| = |\mathcal{L}|,\label{eq:const_1}\\ 
%	&\text{For every $\ell\in\mathcal{L}$: }\exists v\in R: S(v) = 
%	\ell.\label{eq:const_2} 
%\end{align}
\begin{align}
	&\max_{R\subseteq V}\frac{\sum_{e\in E(R)}w(e)}{|R|} \label{eq:opt}\\ 
	&\text{subject to:}\nonumber\\ 
	&~~|R \cap L(\ell)| = 1, \forall \ell \in \mathcal{L} \label{eq:const_2}\\
	&~~|R| = |\mathcal{L}|\label{eq:const_1}
\end{align}~\footnote{Only for the sake of clarity, we have constraints \eqref{eq:const_1} in the problem formulation even though they are redundant given the presence of constraints \eqref{eq:const_2} and the belonging of a gene to exactly one locus.}
The unconstrained version of the above optimization problem (i.e., 
\eqref{eq:opt}) corresponds to the well-known densest subgraph problem 
\cite{goldberg1984finding}, which admits efficient algorithms that will return an optimal 
solution. Adding constraints \eqref{eq:const_1} and \eqref{eq:const_2} as in our case in general makes the problem considerably harder. 
Indeed, variants of the problem similar to the one above are unlikely 
to adimit efficient algorithms that are also optimal \cite{dfsg2020}.

\paragraph{Overview of algorithm.}
In the remainder of this Section, given a weighted graph $G = (V, E, 
w^+)$ with positive weights, we denote by $A$ its (weighted) adjacency 
matrix. Namely, $A$ is a symmetric matrix with non negative entries, 
such that $A_{uv} = 0$ if edge $(u, v)$ does not belong to $E$. 
If $(u, v)\in E$, then $A_{uv}$ equals the corresponding weight $w(u, 
v)$.
The heuristic proposed in this paper proceeds along similar lines as 
the technique proposed  by Anagnostopoulos et 
al.~\cite{dfsg2020} for the extraction of a perfectly balanced dense 
community of nodes from a network. The general idea behind is that, as observed in 
\cite{KV99}, the (unconstrained) densest subgraph problem (i.e., \eqref{eq:opt}) admits a spectral, 
formulation, namely:
\begin{equation}\label{eq:spectral}
	\max_{R\subseteq V}\frac{\sum_{e\in E(R)}w(e)}{|R|} = 
	\max_{\bx\in\{0, 1\}^n}\frac{\bx^TA\bx}{\bx^T\bx}.
\end{equation}
On the other hand, the following relaxation of 
\eqref{eq:spectral}:
\begin{equation*}
	\max_{\|\bx\| = 1}\bx^TA\bx
\end{equation*}
corresponds to computing the main eigenvector of matrix $A$. Moreover, 
as shown in Kannan et al. \cite{KV99}, the presence of a node of 
$G$ in the densest subgraph is correlated with the magnitude of the 
corresponding entry in the main eigenvector of $A$, and a greedy 
technique based on this observation provides a provably approximate 
solution to the densest subgraph problem. Extending this idea, 
Anagnostopoulos et al. \cite{dfsg2020} proposed heuristics to compute dense, fair subgraphs of 
labelled graphs\footnote{In our setting, the label of a node is the locus the 
corresponding gene is associated to.}, based on computing the main 
eigenvector of the following symmetric matrix, called a \emph{fair 
projection matrix}:
\[
	(I - FF^\dagger)A(I - FF^\dagger).
\]
In the setting of \cite{dfsg2020}, a subgraph is \emph{fair} if it 
contains the same number of nodes of each possible label. In the expression above, $F$ is a \emph{fair projection} 
matrix, which enforces the above constraint. We refer the reader to 
\cite[Section 2]{dfsg2020} for full details on $F$ and the arguments 
behind their proposed heuristics.
Unfortunately, the techniques of Anagnostopoulos et al. \cite{dfsg2020} cannot be directly 
applied in our case, since we have the further constraint that the 
dense subgraph we compute should contain exactly one node for each 
label/locus. 
To achieve this, we modified and adapted the heuristic of 
Anagnostopoulos et al. \cite{dfsg2020} to our setting. 
While the pseudocode of our method is described in 
Algorithm~\ref{alg:relmax}, we provide an overview of the algorithm in 
the paragraphs that follow.

\begin{algorithm2e}[h!]
{
\DontPrintSemicolon
 \KwData{$G(V,E, w^+)$, $S \colon V \to \mathcal{L}$, $L \colon \mathcal{L} \to \wp(V)$} 
 \KwResult{$ R \subseteq V,$ a set of nodes of size $|\mathcal{L}|$ with exactly one node per locus.}
 \SetKwProg{Fn}{Function}{:}{}
%\tcp{Data preparation.}
%Remove from $G$ all edges that connect nodes representing genes associated to the same locus.\;
%Remove from $G$ all nodes with no incident edges.\;
\tcp{Solutions generation process.}
$R  \leftarrow \emptyset$; $T = V$.\;
\tcp{$R$ stores the currently best solution.}
\While{$|T| > |\mathcal{L}|$}{
 \tcp{Compute fair projection matrix $F$ for $H = (T, E(T), 
 w_T)$ and $S \colon T \to \mathcal{L}$.}
 Compute the adjacency matrix $A$ of $H$.\;
 Compute $MainEigenVec=MainEigenvector~of~\left(I-FF^\dagger\right)A\left(I-FF^\dagger\right)$.\;
 \tcp{Compute a new solution $R_{current}$ for the problem.}
 $R_{new} \leftarrow \emptyset$.\;
 \ForEach{$\ell \in \mathcal{L}$}{
   $R_{new} \leftarrow R_{new} \cup \argmax_{v \in L\left(\ell\right)} MainEigenVec(v)$.\;
  }
  \tcp{Update the best solution $R$, if it is convenient.}
  \If{$\sum_{(u,v) \in E \mid u \in R_{new} \wedge v \in R_{new} } {w(u, v)} > \sum_{(u,v) \in E \mid u \in R \wedge v \in R } {w(u, v)}$}{
    $R \leftarrow R_{new} $.\;
  }
  %\If{$\frac{\sum_{(u,v) \in E \mid u \in R_{new} \wedge v \in R_{new} } {w(u, v)}}{|R_{new}|} > %\frac{\sum_{(u,v) \in E \mid u \in R \wedge v \in R } {w(u, v)}}{|R|}$}{
  %  $R \leftarrow R_{new} $.\;
  %}
  
  \tcc{Remove from the graph the node with smallest associated main 
  eigenvector's component.}
  $v_{min} = \argmin_{ v \in V \colon |L(S(v))| > 1} MainEigenVec(v)$.\;
  $T \leftarrow T \setminus \{v_{min}\}$
}
\Return $R$\;
}
\caption{\ourmethod Anagnostopoulos et al.~\cite{dfsg2020}}
\label{alg:relmax}
\end{algorithm2e}

The \ourmethod (Algorithm~\ref{alg:relmax}), 
takes in input the 
K-Partite, undirected, and weighted graph resulting from the 
biological preprocessing step described in Section \ref{subse:preproc} 
and the associations between nodes/genes and loci, formally described 
by the functions $S$ and $L$. 
%the function $S \colon V \to \mathcal{L}$, that associates to each node in the graph the locus associated with the gene the node represents, and 
%the function $L \colon \mathcal{L} \to \wp(V)$, that associates to each locus the set of nodes representing the set of genes mapped to the locus.
%The output of the \ourmethod (Algorithm~\ref{alg:relmax}) consists in subset of nodes/genes $R \subseteq V$  
%of size exactly equal to the number of loci we consider, and with exactly one node/gene per locus.
It returns a subgraph that meets constraints $\eqref{eq:const_1}$ and 
\eqref{eq:const_2}, while ideally maximizing \eqref{eq:opt}.
% the ratio between 
%the sum of weights associated with the edges connecting the nodes of the subgraph
%and the number of nodes of the subgraph itself, 
%by keeping 
%a number of nodes exactly equal to the number of loci in input, and exactly one node/gene per locus. 
%The output subgraph is completely defined by the subset of nodes/genes $R \subseteq V$ returned in output by the algorithm. 

%In the data preparation phase of our algorithm (lines 1-2), all   edges that connect genes/nodes associated to the same locus and nodes with no incident edges are removed, since these edges do not contribute to any optimal solution and their presence can only worsen the quality of the solution computed by our algorithm\footnote{This might happen because, as explained before, the problem is generally computationally hard and our heuristic does not compute the optimum in general.}.
%%Due to the fact that the solution must contain exactly one gene per locus and also an high value of the sum of weighs on its edges, 
%these edges and nodes are of no importance for the solution itself and can also potentially reduce the quality of the solution provided by the algorithm.
%After this initial preprocessing, 
The algorithm starts with an iterative process that, in each iteration, computes a feasible solution for the constrained problem \eqref{eq:opt} - \eqref{eq:const_1}, 
and it terminates by providing the feasible solution with the highest density among all the produced ones.
In more detail, in each iteration of the main while loop (line 4 of the 
algorithm), the algorithm processes a weighted subgraph $H = (T, E(T), 
w_T)$ of the original graph $G$ ($G$ itself in the first iteration), 
computing a subset $R_{new}\subseteq R$ of size $|\mathcal{L}|$, so 
that exactly one node/gene in $R_{new}$ is associated to each locus in 
$\mathcal{L}$. To this purpose, modifying the approach proposed in 
\cite{dfsg2020}, it computes the main eigenvector $\bw$ of the fair projection 
matrix corresponding to $H$ and $S: T\rightarrow \mathcal{L}$ (lines 
5-6) and then (lines 7-9), for each locus $\ell$, it 
adds to $R_{new}$ the node/gene associated to $\ell$, whose component 
in $\bw$ is largest. Note that, since $H$ changes across consecutive 
iterations, so do the fair projection matrix and consequently $\bw$. If 
the subgraph induced by $R_{new}$ has higher density than the previous 
best solution, the current best solution $R$ is updated (lines 10-11). 
Finally (lines 12-13), as long as $H$ has more than $|\mathcal{L}|$ 
nodes, the node $v_{\min}$ with the smallest corresponding entry in $\bw$ is 
removed, with $v_{\min}$ chosen from the subset of nodes whose associated loci have at least two associated genes in $H$. 
This amounts to a updating $T$ to $T - \{v_{\min}\}$ and $H$ to 
$\left(T - \{v_{\min}\}, E(T - \{v_{\min}\}), w_{T - 
\{v_{\min}\}}\right)$, after which the next 
iteration of the while loop begins. 
Note that, this way, the number of iterations of the main while loop is 
$|V| - |\mathcal{L}|$.\footnote{It should be noted that the algorithm only explores a 
linear size subset of the exponentially many solutions of the original problem.}

%Each iteration of the ``\texttt{Solutions generation process}'' can be seen as composed by four phases.
%As first phase, the algorithm computes the fair projection matrix $F$ 
%%(according to Anagnostopoulos et al.~\cite{dfsg2020}, from the set of nodes in the graph and the node-locus association)
%, the adjacency matrix of the graph, and the main eigenvector of the resulting matrix from the application of the fair projection to the graph adjacency matrix.
%As shown experimentally in Anagnostopoulos et al.~\cite{dfsg2020}, the application of the fair projection increases the sum of the weights on the solution edges for the same number of nodes.
%Given this main eigenvector, the algorithm generates a feasible solution for the problem by selecting for each locus the node/gene with the higher coordinate value in the eigenvector among all nodes/genes associated with the same locus.
%By simply computing the sum of weights on the edges induced by the current solution, the algorithm decides if it is worth or not to elect the current solution as the best one.
%At the end of each iteration, the algorithm removes from the graph the node with the smallest coordinate value inside the eigenvector, by always keeping at least one node/gene per locus in the graph. This remotion process heuristically peels out from the graph the node in the network that less contributes to the generation of the current solution.
%Finally, among all the generated solutions, only the one with the highest value of the sum of weights on edges is returned in output by the algorithm. 

\section{Results}

\subsection{Experimental Set Up}

To assess the validity of our approach, we compared its performance to FUMA, DEPICT, and MendelVar, which prioritize causal genes at disease risk loci. We further compare our framework with DOMINO, a network-based approach that leverages PPI network topology to find sets of highly connected nodes in the graph involved in the same biological processes. We compared them on two different axes:i) we externally validated various approaches using several data sources and ii) we internally validated them using a newer approach
\medskip

\paragraph{External Validation.} We used \textbf{gseapy} to compute the enriched Reactome and Kegg pathways of each framework's predicted gene set. Furthermore, we considered the precision P defined as:

\begin{equation}
P = \frac{\text{\# of predicted genes }\in\text{ Ground Truth}}{\text{\# of predicted genes}}
\end{equation}

As a Ground Truth, we choose two different test sets. We downloaded drug target association from the DrugHub and we considered as ground truth genes that are drug target of COPD, Bronchospasm and asthma. Then, we downloaded from the Open Target Platform genes involved in respiratory phenotype in mouse model. 
\medskip

\paragraph{Internal Validation.} There is increasing evidence that a set of proteins associated with a given disease do not function in an isolated way. Indeed, these causal proteins interact with each other to form a distinct network module within the universe of (physical) protein-protein interactions (the human protein-protein interactome), representing perturbed, dysfunctional pathways\cite{barabasi2011network, gustafsson2014modules, menche2015uncovering}. Consequently,  starting from independent sets of disease risk loci, a disease-gene prioritization algorithm should prioritize genes involved in similar biological processes. To assess this property, we divided disease-associated SNPs into two groups: Even (E) set, consisting of SNPs located in Even Chromosomes, and Odd (O) set, consisting of SNPs located in the Odd Chromosomes. In this way, we created two independent, balanced inputs(i.e., The chromosome length in one set is similar to the other one)\cite{piovesan2019length}.  Finally, we analyze the sets of genes prioritized using E and O, using a biological similarity measure, named Biological Similarity, and the Network Distance. The former considers common biological processes between predicted gene sets to compute their similarity. While the latter measures the distance between predicted gene sets when projected on a Protein-Protein Interaction network. In the following paragraphs, we formalize these metrics:

\paragraph{Biological Similarity:} We designed the Biological Proximity function  that compares gene sets $\hat{E}$ and $\hat{O}$, predicted respectively starting from E and O,  employing their associated biological pathways. It is  based on the Best Match Average (BMA) rule and employs the Jaccard Index to score gene biological similarity. More formally, given  $\hat{E}$ and $\hat{O}$, the gene sets predicted using respectively Even and Odd Chromosomes, $\varphi(\hat{E}, \hat{O})$ is defined as:

\begin{equation}
    \varphi(\hat{E}, \hat{O}) = \frac{1}{2}  \left(\frac{1}{|\hat{E}|} \sum_{e \in \hat{E}} Sim\left(e, \hat{O}\right) + \frac{1}{|\hat{O}|} \sum_{o \in \hat{O}} Sim\left(o, \hat{E}\right) \right),
\end{equation}

Where $Sim(e,\hat{O})$ score is defined as $\max_{o \in  \hat{O}} {J(e,o)}$ and $J(e,o)$ is the Jaccard Index between biological processes associated respectively with $e$ and $o$.

\paragraph{Network Distance:} we measure the network proximity of modules $\hat{E}$ and $\hat{O}$ as reflected in their gene localizations on Protein Protein Interaction network using the recently introduced separation measure \cite{menche2015uncovering}: 

\begin{equation}
   D_{\hat{E}, \hat{O}} = d_{\hat{E}, \hat{O}} - \frac{ d_{\hat{E}, \hat{O}} + d_{\hat{O}, \hat{O}} }{2}
\end{equation}

which compares the mean shortest distance within the interactome between the predicted genes of each set, $d_{E,E}$ and $d_{0,0}$, to the mean shortest distance $d_{E,O}$ between E–O genes pairs.

\subsection{External Validation: Application to COPD}

\begin{figure*}[t]
\includegraphics[width=\textwidth]{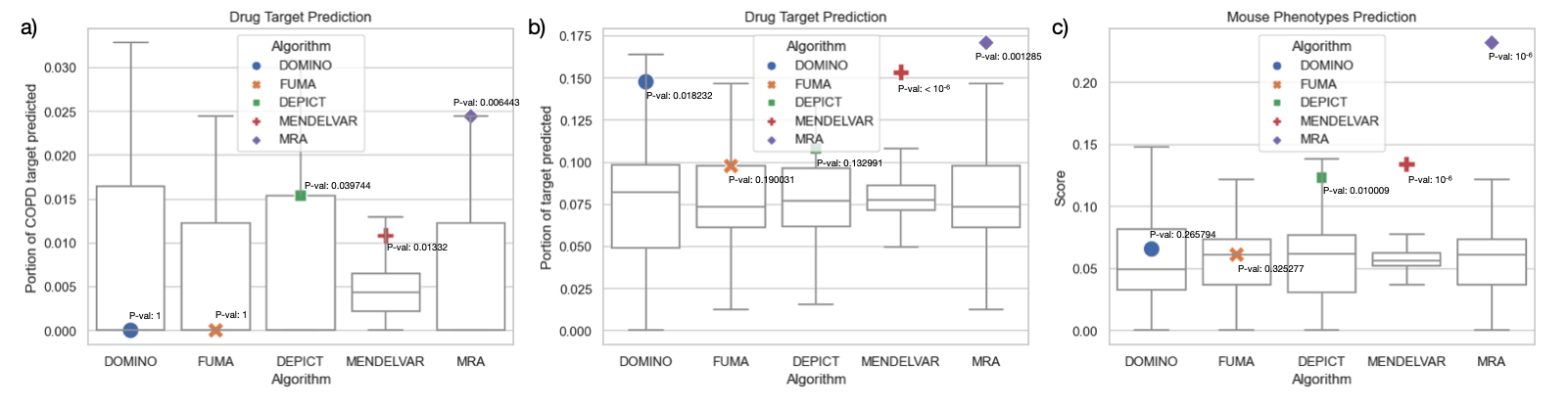}
\caption{\textbf{External Validation}: a) shows the percentage of COPD, asthma, and bronchospasm drug targets predicted by \ourmethod and competing methods. b) shows the percentage of potential drug targets (i.e., FDA approved and In development drugs) predicted by each framework with associated p-values. Finally, c) shows the portion of mutated genes involved in mouse phenotypes (i.e., Respiratory System phenotypes) predicted by each heuristic.  }
\label{fig:external_validation}
\end{figure*}

\subsubsection{Biological Interpretation of prioritized disease genes}
We performed gene-set enrichment analysis to gain functional insight into potentially causal genes assigned to GWAS loci. We compared the statistically significant pathways found by the gene sets associated with genes predicted by \ourmethod with those prioritized by DEPICT, FUMA, MendelVar, and DOMINO. Since \ourmethod is constrained to return one gene per locus, it is tough to have a fair comparison between the algorithm presented in this manuscript. Hence, we designed a different way to find statistically significant Reactome pathways given the set of genes predicted by other heuristics. For each algorithm exception for \ourmethod we grouped their predicted genes by the locus they belong to. Then, we randomly select one gene per locus and compute the enriched pathways using the \textbf{gseapy}, a python library that allowed us to run Enrichr on the python script. We repeated this process several times (i.e., we calculated their statistically significant pathways 1000 times). Finally, we obtained the final p-value of a pathway as an average of all the p-values computed. Table \ref{tab:reactome_pathways} shows the Reactome Pathways that are enriched ($P_{value}$ < 0.05) in at least one of the heuristics analyzed in this manuscript. FUMA, MendelVar, and DEPICT do not find any enrichment in their prioritized gene sets. Instead, \ourmethod and DOMINO returns gene sets that are significantly enriched in the Immune System and the Adaptive Immune System \cite{pouwels2014damps}. Even If \ourmethod prioritizes only one gene per locus, It returns genes statistically significant In Immune System related pathways. Furthermore, It is the only one to find a gene set statistically significant in the Developmental pathways that are thought to play a crucial role in its pathophysiology. \cite{boucherat2016bridging, carlier2020canonical}. 

\begin{figure}
\includegraphics[width=\textwidth]{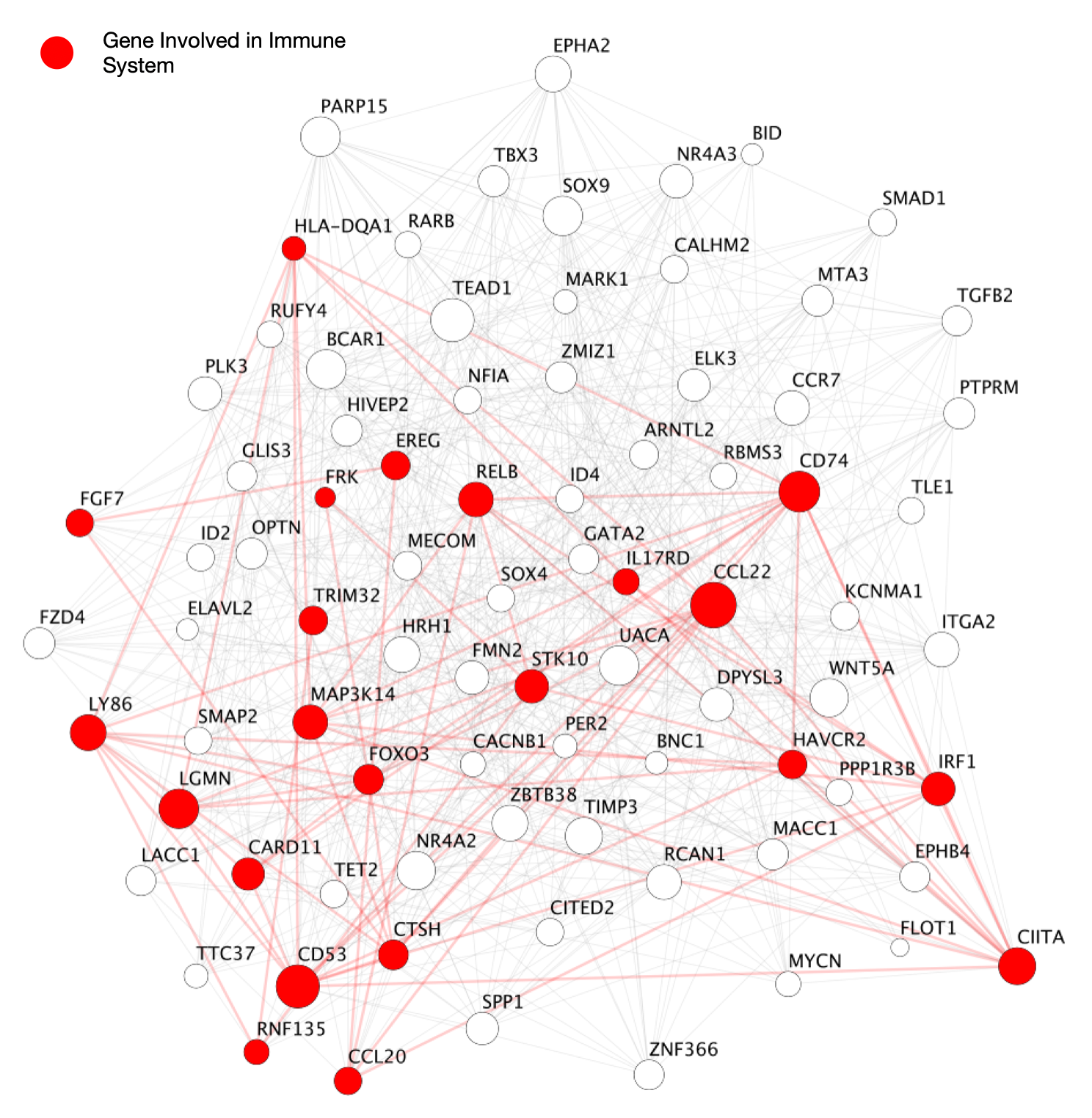}
\caption{Network induced by gene set prioritized by \ourmethod.}
\label{fig:network_solution}
\end{figure}

\medskip
Furthermore, we also considered all set returned by each algorithm to compute Reactome P-values allowing DOMINO, MendelVar, FUMA, and DEPICT to produce more than one gene per locus. Supplementary Table 2 shows the Reactome Enrichment if all prioritized genes are considered. While there seems not to be an improvement in genes predicted by MendelVar, FUMA, and DEPICT, DOMINO enhances the P-values of several Immunological related pathways. One of the possible explanations for this behavior could be related to the input used by DOMINO. Indeed, it is the only one that leverages PPI networks, and it is well known that there is a high correlation between interacting proteins and their shared ontologies. Indeed, Supplementary Figure \ref{fig:domino_and_solution_analysis} a) shows the PPI network induced by the genes prioritized by DOMINO, and each gene has a color that depends on its locus location. Thus, DOMINO prioritizes several biologically related genes located in the same locus, creating a dense module in the PPI network. For instance, It predicts several HLA genes that are well known to be involved in the Immune System related pathways.

\subsubsection{Identification of drug targets}

\noindent We examined the 82 gene-locus associations returned by the \textbf{\ourmethod} and we compared them with the gene-locus associations predicted by the other heuristics. First, we compared the outputs of all algorithms by considering, as ground truth, only genes that are targets of ``launched drugs'' (i.e., approved by FDA) that are used to treat asthma, bronchospasm, bronchitis or chronic obstructive pulmonary disease. We found that  \textbf{\ourmethod} is able to predict 2 drug targets, namely, \textit{KCNMA1} and \textit{HRH1} and  DEPICT only predicts one:  \textit{ADRB2}. MendelVar prioritizes the highest number of COPD related drug targets: TNF, CHRM3, SLC6A4, HRH1, KCNMA1. Instead, FUMA and DOMINO does not prioritize any COPD related drug targets. We should notice that DEPICT, DOMINO and MendelVar in general associate a list of genes to each locus. For this reason, we compared each the portion of drug target found by each algorithm (Precision) with its random expectation. Figure \ref{fig:external_validation} b) shows for each algorithm the portion of prioritiized genes that are targets of COPD, asthma, and bronchospasm FDA approved Drugs we downloaded from The Drug Hub.
 
\noindent We then broadened our scope, considering the entire drug target association data set. We found that 14 genes (\textit{EPHA2, PLK3, ITGA2, CACNB1, HRH1, PARP15, TIMP3, MARK1, RARB, FRK, EPHB4, STK10, WNT5A} and \textit{KCNMA1}) are drug targets of several \textbf{in Development} or \textbf{Launched} drugs (see Supp. Table 2 for more information \footnote{\url{https://docs.google.com/spreadsheets/d/1cWy2fGU9BY6bZFzlr56fmz74z1o4kA3TZE80fXDYxmc}}). However, MendelVar is the algorithm that retrieves the largest number of drug targets. As we discussed previously, to make the comparison fair, we analyzed the precision of each algorithm (i.e., the portion of predicted genes that are target of any Launch or In Development Drugs). If we consider the precision, our heuristic has the highest value, followed by MendelVar and DEPICT. it is worth noticing that comparing each framework with its own random distribution, the onlyones that obtains statistically significant results are MendelVar ($P_{value} < 10^{-6}$) and \ourmethod ($P_{value}  0.001285 $).

\subsubsection{Identification of Genes involved in Mouse Respiratory Phenotypes}

\noindent The use of non-human species to understand regular and pathobiology and to create models of human diseases tractable to the experimental investigation has been a dominant and successful paradigm in the biomedical sciences for many years\cite{schofield2012mouse,rosenthal2007mouse}. Consequently, we externally validate gene sets prioritized by the frameworks considered in this manuscript using the Open Target Platform, a resource of mouse phenotypes and mapped genes. We download all genes involved in "respiratory phenotypes" from the Open Target Platform to derive a validation set. Then, we computed the precision (i.e., the fraction of prioritized genes in the validation set), and we compared each algorithm's accuracy with their random distribution. Figure \ref{fig:external_validation} d) shows the performance of each framework compared to its random expectation. Only three framework obtains a statistically significant accuracy: \ourmethod has the best prediction accuracy that is also statistically significant (i.e., $P_{value} < 10^{-6}$), MendelVar, that predicts the largest number of genes mapped in mouse phenotypes, and DEPICT. The list of genes predicted and their mouse phenotypes is visible on Supplementary Table 3.

\begin{figure*}[t]
\centering
\includegraphics[width=\textwidth]{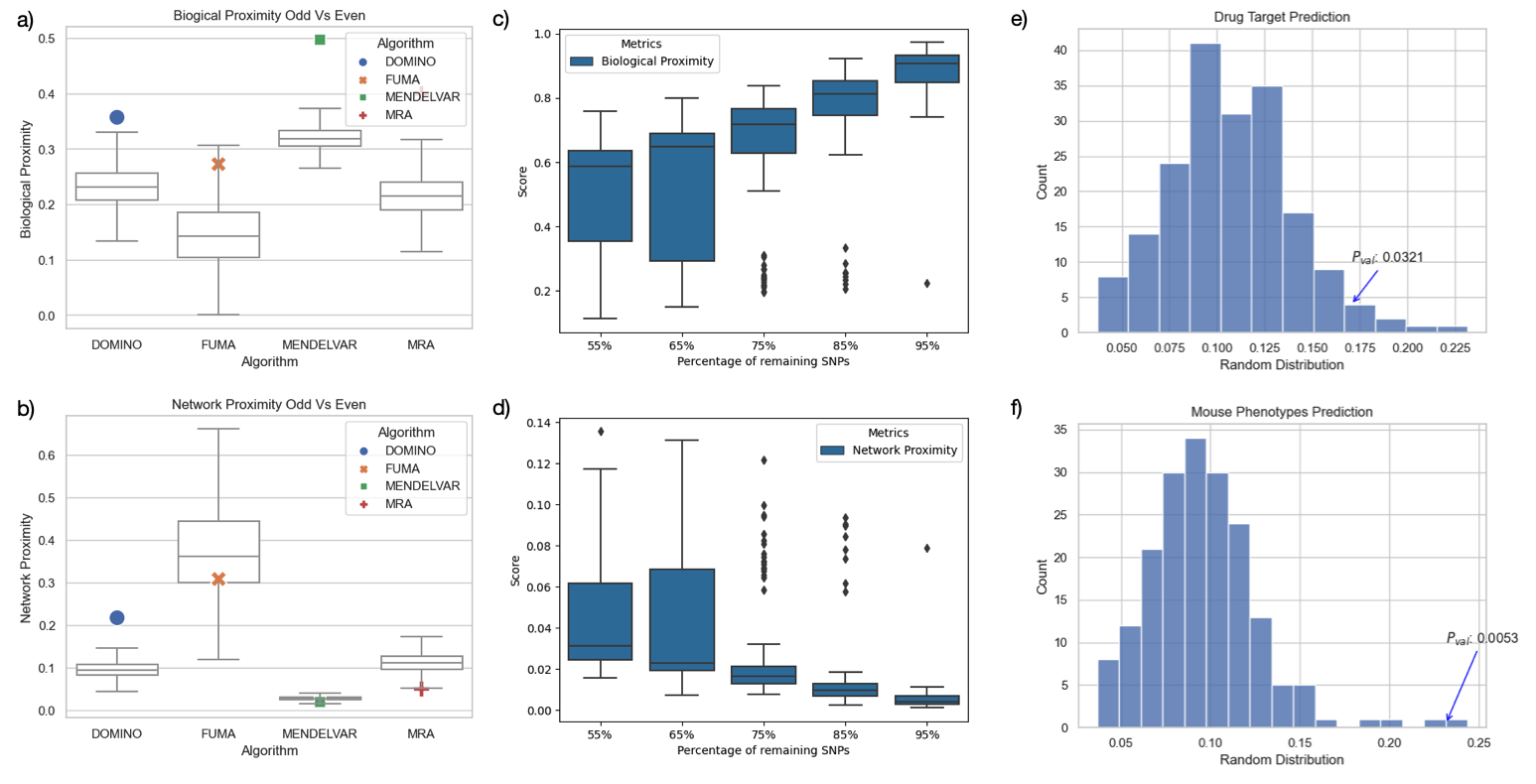}
\caption{\textbf{Internal Validation}: a) and b) show the biological similarity and network distance between predicted gene sets prioritized using respectively even and odd chromosomes. Since the number of variants in the two inputs is small DEPICT did not return any statistically significant gene sets. c) and d) show the behaviour of Biological Similarity and network distance on the gene sets predicted by \ourmethod when the input is affected by missing information. Finally, e) and f) show how random information can affect the prediction of Drug Targets and gene involved in Respiratory system (Mouse phenotypes)}
\label{fig:sensitivity_analysis}
\end{figure*}

\subsection{Cross Study Validation: Application to Breast Cancer}

To further understand the ability of our framework to predict robust, biologically meaningful results, we downloaded several studies on Breast Cancer from The GWAS Catalog and compared the gene sets returned by \ourmethod. Here we present the results using BC studies with enough information (i.e., a good amount of SNPs found in each study). First, we plot the heatmap representing the similarity between pairs of studies in terms of statistically significant SNPs. Indeed, Supplementary Figure \ref{fig:cross_validation} a) shows the Jaccard Index between enriched SNPs. As we can see, a bunch of studies shares more than 30\% of SNPs, and as a consequence, the Jaccard Index of their predicted gene set is very similar, as shown in Supplementary Figure \ref{fig:cross_validation} b). This result depends on the fact that a good part of these studies (i.e., GCST001937, GCST003842, GCST90090980, GCST003845, GCST004950, GCST007236) is derived from the same population. However, one study (i.e., GCST90011804), that does not share common SNPs with the others, share part of its predicted gene set (i.e., IJ > 0.16). Indeed, GCST90011804 uses a different cohort but nevertheless it can return genes that are involved in similar biological processes as show in Supplementary Figure \ref{fig:cross_validation} c). Furthermore, all predicted gene sets are closer in the Protein-Protein interaction network as shown in \ref{fig:cross_validation} d), where the network distance varies between 0.0 and 0.07.

\subsection{Internal Validation}

\subsubsection{Odd vs. Even }
\label{sub:odd_vs_even}

Figure \ref{fig:sensitivity_analysis} a) and b) show for each algorithm the biological and network proximities between prioritized gene sets (i.e., odd and even split) and compares them with a random distribution. We calculated the biological proximity using ontologies (biological processes) downloaded from Gene Ontology Consortium. Furthermore, to avoid possible correlations between network proximity and the genes prioritized by DOMINO, we change the PPI network in which the network proximity is computed. Although DOMINO leverages the protein-protein interaction network, when we divide the seed set into two different sets, the resulting prioritized gene sets are very far from each other in terms of network proximity. Indeed, the associated empirical p-value is circa 1. Interestingly, all frameworks return  gene sets (i.e.,$\hat{E}$ and $\hat{O}$) that are biologically similar compared to random expectation (i.e., empirical p-value < 0.05). However, when we compute the network proximity of odd and even prioritized sets, only MendelVar and \ourmethod return sets of genes located in the same protein-protein interaction network area. Furthermore, DEPICT cannot prioritize any genes since the input information (31 even loci and 50 odd risk loci) is not enough to return statistically significant annotations.

\subsubsection{The General Approach}
In section \ref{sub:odd_vs_even}, we demonstrated the ability of the \ourmethod to find biologically related gene sets  starting from independent inputs ( i.e. the 2 sets of SNPs associated respectively to Odd and Even Chromosomes). Here, we discuss a more general approach to show the ability of our framework to prioritize biologically similar gene sets when the information is missing.

\medskip\noindent
To introduce noise (i.e missing information) on our dataset, consisting of 82 Statistically significant SNPs associated with Chronic Obstructive Pulmonary Disease, we first removed a portion $i$ of them uniformly at random and we repeated this process to get a fold of 100 inputs. Secondly, we execute the  \ourmethod on each noisy input to prioritize a gene set. Finally, we computed the Biological and Network proximity between each prioritized gene set and the gene set induced by all the 82 enriched SNPs. To better understand how missing information affects our predictions, we repeated the overall experiment removing  a portion of data in the interval [5\%, 50\%].

\noindent Figure \ref{fig:sensitivity_analysis} c) and d) show how missing data affect the similarity between a noisy dataset and the original one. a)  and b) show how much biological and network proximity are affected by missing information. Each boxplot shows the distribution of biological and network proximities when a percentage of SNPs is removed from the algorithm's inputs. When 50\% of SNPs are removed, the output is noisy and very distant from the original solution in terms of biological similarity and network proximity. Furthermore, when the missing information is reduced (i.e., 75\% to 95\% of SNPs are the input of \ourmethod), we can see how biological proximity converges to 1 and network proximity to 0. Thus, \ourmethod can still find a suitable solution if we remove at most 25\% percent of our knowledge from the input. Unfortunately, we can do this experiment only on \ourmethod since FUMA and MENDELVAR are web applications that we cannot run locally.

\subsubsection{Randomization and Bias }

Unlike other network-based approaches, \ourmethod weights the co-regulation network using Gene Ontologies as described in section 3.2.1. Since manually curated information creates a bias in the co-regulation network, we analyze if the set of genes predicted is driven exclusively by the bias induced by GO or if it depends on the risk locus-gene associations. Consequently, we create 100 random risk locus-gene bipartite graphs without changing the number of associations of each risk locus. Thus, the co-regulation network induced by a randomized bipartite network is always different (i.e., links between genes associated with the same risk locus are removed). For each generated co-regulation network, we ran the \ourmethod and computed the portion of predicted drug targets and the percentage of genes involved in mouse phenotypes. Figure \ref{fig:sensitivity_analysis} e) shows the distribution of the percentage of drug targets predicted using random networks compared with the score predicted using not randomized data ($P_{val} \simeq 0.03$). Finally, figure \ref{fig:sensitivity_analysis} f) shows the distribution of the portion of genes involved in mouse phenotypes predicted using random networks and compared with the gene predicted using the original bipartite graph ($P_{val} \simeq 0.0053$). As expected, the solution returned using original data is more biologically meaningful than those returned using a randomized risk locus-gene association dataset.

\section{Conclusion}

This article proposes a novel method, \ourmethod, to identify the single most likely causal gene at GWAS risk loci. \ourmethod associates each causal variant with the most probable causal gene identifying the set of genes (i.e., one for each casual variant) with the highest density in a weighted gene co-regulation network. Unlike previous network-based disease genes prioritization algorithms such as dmGWAS or DOMINO, \ourmethod considers only a subgraph consisting of edges connecting nodes that do not overlap the same locus. Then, It searches for a subset of nodes, one per each locus, that maximizes the total number of connections. Furthermore, instead of leveraging PPI topology, this framework exploits the structure of a CO-regulation network where a pair of genes is connected if they are co-regulated by the same transcription factors.
\newline

\noindent We biologically validated our framework on COPD, a complex disease with a set of 82 risk loci previously discovered through GWAS. Since there is no ground truth about the proper association between COPD risk loci and the causal gene at each risk locus, we compared the \ourmethod with the existing methods on three different axes: i) statistically significant Reactome and KEGG Pathways identified by each prioritized set gene of putative causal genes, ii) the portion of drug targets found by each heuristics, and iii) the portion of genes that have been previously identified to be involved in respiratory system mouse phenotypes.
\newline

\noindent Although our gene prioritization results for COPD GWAS were plausible based on prior COPD research, there is no ground truth for the causal genes and biologic networks underlying COPD pathogenesis. Thus, we did not have a way to compute and compare the specificity and sensitivity of each gene prioritization algorithm for COPD. Consequently, we validated the consistency of \ourmethod in the face of data incompleteness. To reduce the signal present in the COPD GWAS study, We divided the risk loci into two independent sets containing risk loci in Even and Odd chromosomes. We ran \ourmethod and other GWAS prioritization algorithms considering these sets as inputs, and we analyzed the results using two different approaches. First, we compared the prioritized gene sets in terms of biological similarity, comparing the set of biological processes they are involved in. Secondly, we compare the outputs based on their distance in the PPI network. Except for DEPICT, we found that all the algorithm shows similar results and predict meaningful results even if the input is subjected to data incompleteness. Furthermore, we repeated this experiment on Breast Cancer disease to see if the performances were consistent. We downloaded several GWAS studies, and we analyzed the biological and topological similarities of the prioritized gene sets.

Finally, we generalized the internal validation, randomly removing a portion of risk loci from the input and compared an incomplete solution with that one in which all the GWAS data is considered. We found that \ourmethod is resilient to missing information and return a meaningful solution even if the 25\% of data is removed. Unfortunately, since MendelVar and FUMA are web servers, we did not have the chance to repeat this experiment on them.

This framework showed promising results when compared with the state-of-the-art. However, we can still improve this heuristic to return more reliable predictions. Firstly, it is relatively easy to adapt the framework to work with other data types, such as e-QTLs, to help the heuristic choose the optimal casual gene for a given locus. Secondly, the method can leverage only one network. Still, we may consider modeling the problem on a multi-layer graph, considering other networks such as PPI and CO-Expression.

In summary, we introduce the \ourmethod for jointly prioritizing putative causal genes across all GWAS loci for a trait or disease and selecting one biologically similar potential causal gene at each genetic risk locus. The \ourmethod outperforms related gene prioritization methods in capturing biologically-relevant information measured by drug targets and mouse phenotypes while remaining robust to incomplete and noisy GWAS data. 

\section{Acknowledgement}
This work is partially supported by the ERC Advanced Grant 788893 AMDROMA ``Algorithmic and Mechanism Design Research in Online Markets'', the EC H2020RIA project ``SoBigData++'' (871042), and the MIUR PRIN project ALGADIMAR ``Algorithms, Games, and Digital Markets''. BDH is supported by NIH K08 HL136928, U01 HL089856, R01 HL135142, R01 HL139634, and R01 HL147148; an Alpha-1 Foundation Research Grant; and receives grant support from Bayer.

\bibliographystyle{alpha}
\bibliography{bibliography}

\appendix

\newpage
\setcounter{figure}{0}
\renewcommand{\figurename}{Supplementary Fig.}

\begin{appendices}
\section{Results}\label{se:apx_data}
\subsection{Cross Validation}
\begin{figure*}[pht]
\centering
\includegraphics[width=\textwidth]{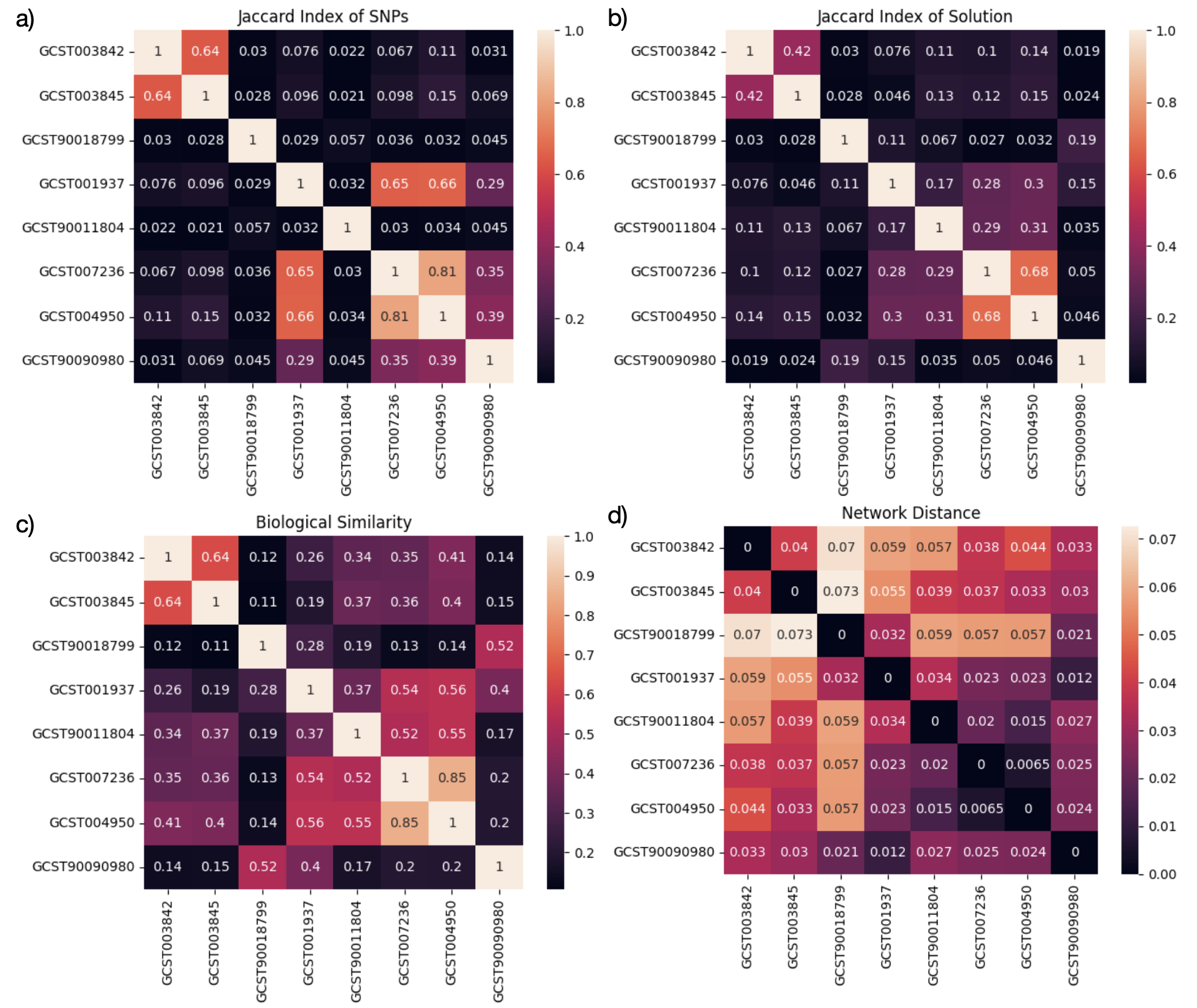}
\caption{Comparison between gene sets prioritized using different set of SNPs downloaded from The GWAS Catalog. a) shows the Jaccard Index of sets of SNPs between each pair of studies. b) shows the Jaccard Index of prioritized gene sets between each pair of studies. c) and d) shows the Biological Similarity and the Network Distance between each pair of predicted gene sets.}
\label{fig:cross_validation}
\end{figure*}
\subsection{Analysis of predicted Gene Sets}

\begin{table*}[!htp]\centering
\caption{Algorithm Comparison on Statistically significant Reactome and KEGG Pathways. The table shows Pathways that are enriched in at least one heuristics (i.e., $P_{value} < 0.05$). We use $-Log_{10}$ scale to convert each $P_{value}$.}
\label{tab:reactome_pathways}
\scriptsize
\begin{tabular}{p{2.0cm}p{6.0cm}p{1.0cm}p{1.0cm}p{1.0cm}p{1.3cm}p{1.3cm}}
\textbf{Dataset}&\textbf{Pathways} &\textbf{DOMINO} &\textbf{FUMA} &\textbf{DEPICT} &\textbf{MENDELVAR} &\textbf{MRA} \\
 \multirow{16}{*}{\textbf{Reactome}}&\textbf{Immune System} &3.91 &0.04 &0.16 &0.14 &2.23 \\
&\textbf{Cytokine Signaling in Immune system} &4.21 &0.11 &0.59 &0.28 &1.49 \\
&\textbf{Adaptive Immune System} &1.38 &0.07 &0.30 &0.18 &1.49 \\
&\textbf{MHC class II antigen presentation} &1.82 &0.00 &0.00 &1.10 &1.49 \\
&\textbf{Innate Immune System} &0.91 &0.32 &0.33 &0.18 &2.23 \\
&\textbf{Hemostasis} &0.67 &0.04 &0.99 &0.23 &1.47 \\
&\textbf{Axon guidance} &0.40 &0.38 &0.89 &0.28 &1.49 \\
&\textbf{Developmental Biology} &0.27 &0.40 &0.89 &0.22 &1.49 \\
&\textbf{Nuclear Receptor transcription pathway} &0.00 &0.50 &0.90 &0.55 &1.49 \\
&\textbf{TCF dependent signaling in response to WNT} &0.00 &0.21 &0.96 &0.55 &1.49 \\
&\textbf{Deactivation of the beta-catenin transactivating complex} &0.00 &0.00 &0.00 &0.50 &1.49 \\
&\textbf{WNT5A-dependent internalization of FZD4} &0.00 &0.00 &0.00 &0.52 &1.49 \\
&\textbf{Dectin-1 mediated noncanonical NF-kB signaling} &0.00 &0.00 &0.00 &0.78 &1.47 \\
&\textbf{CLEC7A (Dectin-1) signaling} &0.00 &0.00 &0.00 &0.72 &1.49 \\
&\textbf{C-type lectin receptors (CLRs)} &0.00 &0.00 &0.00 &0.64 &1.49 \\
&\textbf{Chemokine receptors bind chemokines} &0.00 &0.00 &0.00 &1.20 &1.49 \\
\multirow{14}{*}{\textbf{KEGG}}&\textbf{Antigen processing and presentation} &3.8666 &0.2606 &0.0000 &0.7843 &2.1558 \\
&\textbf{Tuberculosis} &1.3978 &0.0000 &0.4267 &0.4671 &2.5623 \\
&\textbf{C-type lectin receptor signaling pathway} &0.7868 &0.0000 &0.0000 &0.8774 &2.6169 \\
&\textbf{Hippo signaling pathway} &0.7113 &0.0000 &0.7527 &0.3770 &2.6169 \\
&\textbf{MAPK signaling pathway} &0.8027 &0.2606 &2.0726 &0.4271 &2.7119 \\
&\textbf{PI3K-Akt signaling pathway} &0.7529 &0.2606 &0.9459 &0.3876 &1.4735 \\
&\textbf{Gastric cancer} &1.8639 &0.2606 &2.2511 &0.4135 &2.1287 \\
&\textbf{Proteoglycans in cancer} &1.7971 &0.0000 &1.4500 &0.3517 &1.7392 \\
&\textbf{Pathways in cancer} &1.5196 &0.2606 &2.7269 &0.3510 &1.7392 \\
&\textbf{TGF-beta signaling pathway} &0.0000 &0.2606 &2.1576 &0.4636 &1.9723 \\
&\textbf{Signaling pathways regulating pluripotency of stem cells} &0.0000 &0.2434 &0.5587 &0.3970 &2.7119 \\
&\textbf{Chemokine signaling pathway} &0.0000 &0.2062 &0.0000 &0.5249 &1.7858 \\
&\textbf{Mitophagy} &0.0000 &0.0000 &0.0000 &0.5385 &1.5257 \\
&\textbf{Transcriptional misregulation in cancer} &0.0000 &0.0000 &0.0000 &0.6232 &1.7858 \\
\end{tabular}
\end{table*}

\begin{figure*}[pht]
\centering
\includegraphics[width=\textwidth]{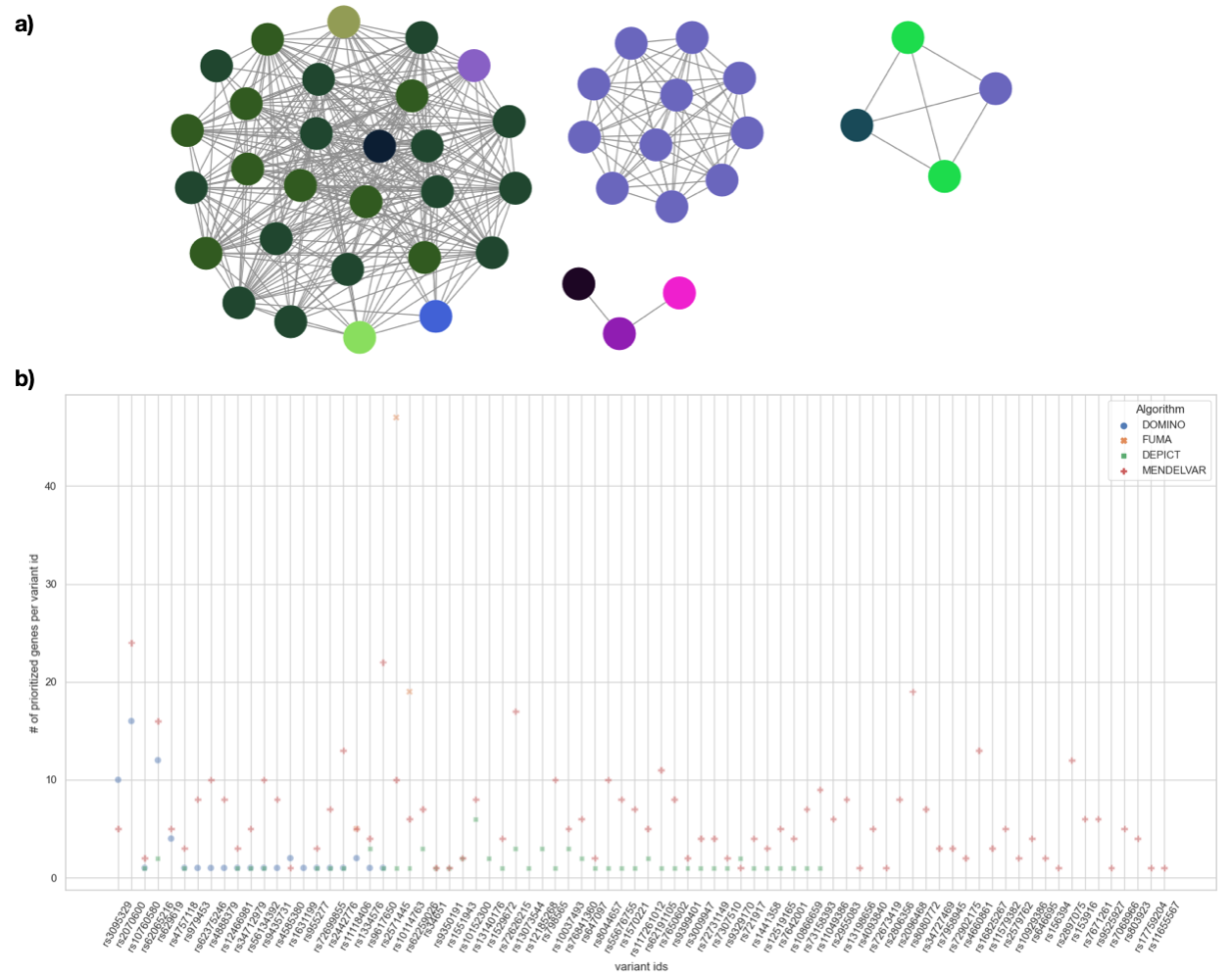}
\caption{\textbf{Predicted Gene Set Comparison}: a) Shows the PPI sub-network induced by the gene set prioritized by DOMINO. Each node is colored based on the locus it belongs to. b) shows the number of gene predicted by variant id. Since we predict only one gene per locus, we removed \ourmethod from this comparison. }
\label{fig:domino_and_solution_analysis}
\end{figure*}
\end{appendices}

\end{document}